\documentclass[aps,prl,amsmath,twocolumn,superscriptaddress]{revtex4}
\usepackage{graphicx}
\usepackage{bm}
\usepackage{color}

\begin{document}

\title{Valley contrasting chiral phonons in monolayer hexagonal lattices}
\author{Lifa~Zhang}
\affiliation{Department of Physics, The University of Texas at Austin, Austin, Texas 78712, USA}

\author{Qian~Niu}
\affiliation{Department of Physics, The University of Texas at Austin, Austin, Texas 78712, USA}
 \affiliation{International Center for Quantum Materials, Peking University, Beijing 100871, China}

\begin{abstract}
In monolayer hexagonal lattices, two inequivalent valleys appear in the Brillouin zone. With inversion symmetry breaking, we find chiral phonons with valley  contrasting circular polarization and ionic magnetic moment. At valley centers, there is a three-fold rotational symmetry endowing phonons with a quantized pseudo angular momentum, which includes spin and orbital parts. From conservation of the pseudo angular momentum, crystal momentum and energy, selection rules in intervalley scattering of electrons by phonons are obtained. The chiral valley phonons are verified and the selection rules are predicted in monolayer Molybdenum disulfide.  Due to valley contrasting phonon Berry curvature, one can also detect a valley phonon Hall effect.  The valley-contrasting chiral phonon, together with phonon circular polarization, ionic magnetic moment, phonon pseudo angular momentum, valley phonon Hall effect, will form the basis for valley-based electronics and phononics applications in the future.
\end{abstract}

\maketitle
Due to inversion symmetry breaking, valley-contrasting electronic physics has been proposed by Xiao \emph{et al}. in 2007 \cite{xiao07}, and followed by valley-dependent optoelectronics \cite{yao08} and coupled spin and valley physics in monolayers of group-VI dichalcogenides AB$_2$ (A$=$ Mo, W; B$=$S, Se)) lattices \cite{xiao12}. Besides charge and spin, separated valleys in momentum space constitute another discrete degrees of freedom for electrons with long relaxation time, which leads to emergence of valleytronics,  such as valley polarization and valley coherence on  transition-metal dichalcogenides monolayer AB$_2$  materials \cite{cao12,wang12,mak12,zeng12,wu13,jones13,xu14}. The valley electron interband scattering involves a polarized photoexcitation and photoluminescence; however, the intervalley electron scattering will involve $\bm K$-valley phonons \cite{zeng12}. Given the fact that electrons have definite chirality at valleys, a natural question then arises: whether do valley phonons have chirality and how does the chirality play a role in electronic intervalley scattering?
\begin{figure}[th]
\includegraphics[width=3.4 in, angle=0]{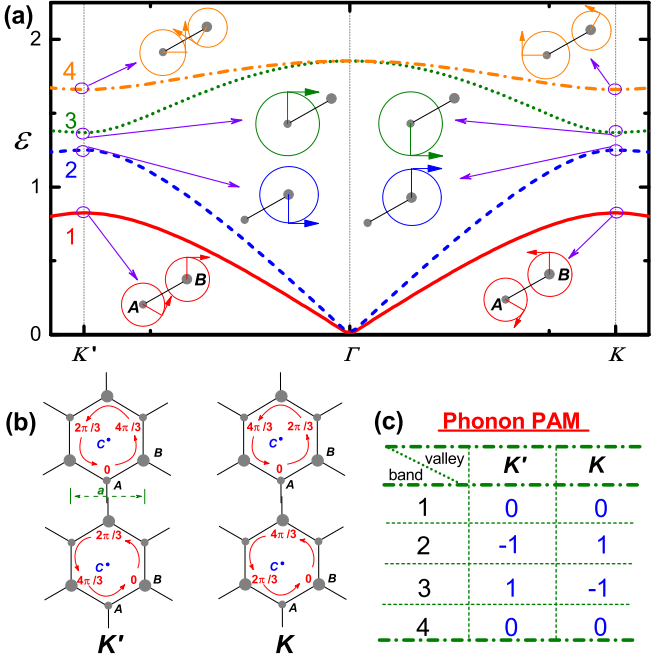}%
\caption{\label{fig1} \textbf{Valley phonons in a honeycomb AB lattice.} (a) Phonon dispersion relation of a honeycomb AB lattice. The insets show phonon vibrations for sublattices A and B in one unit cell at $\bm{K'}$ ($k_x=-\frac{4\pi}{3a},k_y=0$) and $\bm{K}$ ($k_x=\frac{4\pi}{3a},k_y=0$),  numbers 1 to 4 denote four bands. The radii of circles denote vibration amplitudes, phase and rotation direction are included.  (b) Phase correlation of phonon non-local part for sublattice A (upper two panels) and sublattice B (lower two panels) at $\bm(K')$ (left) and $\bm{K}$ (right). (c) Phonon pseudo angular momentum for bands 1 to 4 at valleys $\bm{K'}$ and $\bm{K}$.  Here, the longitudinal spring constant $K_L=1$, the transverse one $K_T=0.25$, and  $m_A=1$, $m_B=1.2$.  The primitive vectors are $(a, 0)$ and $(a/2, \sqrt{3}a/2)$.}
\end{figure}

In this Letter, we observe phonon chirality at two valleys in honeycomb systems with broken inversion symmetry.  By introducing isotopic doping \cite{chen12} or  staggered sublattice potential \cite{zhou07} in graphene,  the inversion symmetry can be broken.  The monolayer hexagonal Boron Nitride which can be successfully grown over a wide area \cite{kim13},  can also be expected to present valley chiral phonons. Besides these kinds of monolayer honeycomb AB lattices, monolayer hexagonal AB$_2$ lattices which have been widely investigated in valleytronics can also serve as paradigms for applications of $\bm K$-valley chiral phonons.  The chirality of phonons at $\bm K$-valleys not only decides the selection rules in electronic intervalley scattering but also can endow phononics with other potential effects, e.g., valley phonon Berry curvature and valley phonon Hall effect. And near $\bm K$-valley centers phonon has extremes in dispersion and thus a large density of states. Therefore chiral valley phonons will play an important role in valleytronics, especially in intervalley scattering of electrons or holes.

\vspace{6pt}
\textbf {Chirality of $\bm K$-valley phonons.} To study $\bm K$ valley phonons, we first focus on a two-dimensional honeycomb lattice model, where each unit cell has two sublattices A and B. The honeycomb AB lattice can serve as a simplified model to demonstrate general features of chiral phonons and can be generalized to other honeycomb materials with broken inversion symmetry.  The dynamical matrix is obtained by considering nearest neighbor interaction. The spatial inversion symmetry is broken if the two sublattices A and B have different mass when a gap is open between the acoustic bands and optical bands as shown in the $\bm K$ and $\bm K'$ points in Fig.~\ref{fig1}(a). From calculated eigenvectors, we can plot sublattice vibrations at valleys as shown in the insets of Fig.~\ref{fig1}(a).  At valleys, all the vibrations are circularly polarized. For valley phonon modes of band 1 and 4, both sublattices do opposite circular motion; for band 2 and 3 while one sublattice is still the other sublattice does a circular motion.  From valley $\bm K$ to $\bm K'$, all the circular motions will change to opposite directions. In latter discussion, we will focus on the $\bm K$ point while results at $\bm K'$ are achieved by time reversal symmetry.

The phonon chirality can be characterized by polarization of phonon, which comes from the circular polarization of sublattices. To consider the polarization along $z$ direction, we look at the phonon eigenvectors {\small $\epsilon=\left( x_1 \; y_1 \; x_2 \; y_2  \right)^T $ }(here we take a two-sublattice unit cell as an example, for a general case see Supplementary information Sec. I).  By defining a new basis where one sublattice has right-handed or left-handed circular polarization as {\small{
$\left| {R_1 } \right\rangle  \equiv \frac{1}{{\sqrt 2 }}\left( 1 \; i \; 0 \; 0  \right)^T $; $ \left| {L_1 } \right\rangle  \equiv \frac{1}{{\sqrt 2 }}\left( 1 \; -i\; 0\;  0 \right)^T $; $ \left| {R_2 } \right\rangle  \equiv \frac{1}{{\sqrt 2 }}\left(0 \;  0 \; 1 \; i  \right)^T $; $\left| {L_2 } \right\rangle  \equiv \frac{1}{{\sqrt 2 }}\left( 0 \;  0 \; 1 \; -i \right)^T $,}} the phonon eigenvector $\epsilon$ can be represented as
\begin{equation}\label{eq_eps}
\epsilon  = \sum\limits_{\alpha = 1}^n {\epsilon _{R_\alpha } \left| {R_\alpha } \right\rangle  + \epsilon _{L_\alpha } \left| {L_\alpha } \right\rangle},
\end{equation}
where $\epsilon _{R_\alpha }  = \left\langle {R_\alpha } \right|\left. \epsilon  \right\rangle  = \frac{1}{{\sqrt 2 }}(x_\alpha  - iy_\alpha ),\;\;\epsilon _{L_\alpha }  = \left\langle {L_\alpha } \right|\left. \epsilon  \right\rangle  = \frac{1}{{\sqrt 2 }}(x_\alpha  + i y_\alpha )$. Then the operator for phonon circular polarization along $z$ direction can be defined as
\begin{equation}
\hat S^{z} \equiv \sum\limits_{\alpha = 1}^n {\left( {\left| {R_\alpha } \right\rangle \left\langle {R_\alpha } \right| - \left| {L_\alpha } \right\rangle \left\langle {L_\alpha } \right|} \right)} ,
\end{equation}
the phonon circular polarization equals to
\begin{equation} \label{eq_spinang}
s_{\rm ph}^{z}= \epsilon ^\dag  \hat S^{z}  \epsilon \hbar= \sum\limits_{\alpha = 1}^n {\left( {\left| {\left. {\epsilon _{R\alpha} } \right|} \right.^2  - \left| {\left. {\epsilon _{L\alpha} } \right|} \right.^2 } \right) \hbar},
\end{equation}
here $n=2$ for two-sublattice unit cells. The phonon circular polarization can have a value between $\pm\hbar$ since $\sum\limits_\alpha {\left| {\left. {\epsilon _{R_\alpha } } \right|} \right.^2  + \left| {\left. {\epsilon _{L_\alpha } } \right|} \right.^2 }  = 1$.  The $s_{\rm ph}^{z}$ has the same form with that of phonon  angular momentum  $j_{\bm k, \sigma}^z$ along $z$ direction  (see Supplementary information Sec. I). The phonon circular polarization comes from those of sublattices in a unit cell. Each sublattice can have a circular polarization $s_\alpha^z$ which equals to $\epsilon ^\dag  \hat S_\alpha^z \epsilon \hbar$ with $\hat S_\alpha^z=\left| {R_\alpha } \right\rangle \left\langle {R_\alpha } \right| - \left| {L_\alpha } \right\rangle \left\langle {L_\alpha } \right|$. In Fig.~\ref{fig1}(a), at valley $\bm K$, $s_A^z=0,\,s_B^z=-\hbar$ for band 2 while $s_A^z=\hbar,\,s_B^z=0$ for band 3, where the phonon circular polarization happens to be quantized; for band 1 and 4, sublattice A (B) is left (right) circularly polarized with $|s_A^z|<1$ ($|s_B^z|<1$).  Therefore, phonon circular polarization $s_{\rm ph}^{z}$ at valleys can be nonzero. If the sublattices have nonzero effective charges $e_\alpha$, one also can expect nonzero ionic magnetic moment $M_z=\sum\limits_{\alpha = 1}^n {s_\alpha^z\,e_\alpha/(2m_\alpha)}$ at valleys.  By introducing a staggered sublattice onsite potential, the valley phonon angular momentum can also be observed (see Supplementary information Sec. II).

\vspace{6pt}
\noindent \textbf {Phonon pseudo angular momentum at valleys.} In a honeycomb lattice, at high symmetry points $\bm \Gamma, \bm K, \bm K'$  phonons are invariant under a 3-fold discrete rotation about the direction ($z$) perpendicular to the lattice plane. Under the rotation, one can obtain $\Re(\frac{2\pi}{3},z) u_{\bm k}= e^{-i\frac{2\pi}{3}l^{\bm k}_{\rm ph}}u_{\bm k} $, where $l^{\bm k}_{\rm ph}$ is defined as the pseudo angular momentum of phonon with wave function $u_{\bm k}$. The phase correlation of the wave function comes from two parts, one is from local (intra-cell) part $\epsilon_{\bm k, \sigma}$, another is from the non-local (inter-cell) part $e^{i{\bm R}_l\cdot{\bm k}}$. Thus under a 3-fold rotation, at valleys one can obtain  spin pseudo angular momentum $l^s$ for the local part and orbital pseudo angular momentum $l^o$ for the non-local part. The math of pseudo angular momentum under 3-fold rotator is $1+1=-1$ since the eigenvalues of rotation are $\pm1,0$.

Under a 3-fold rotation, both circular polarization $\left| {R_\alpha } \right\rangle $ and $ \left| {L_\alpha } \right\rangle $ are eigenstates of the operator $\Re(\frac{2\pi}{3},z)$ with pseudo angular momentum $l^s_R =1$ and $l^s_L=-1$ respectively. Therefore, in Fig.~\ref{fig1}(a), at valley $\bm K$, $l^s_A=-1,\,l^s_B=1$ for band 1 and 4, $l^s_B=-1$ for band 2 and $l^s_A=1$ for band 3.   Since a 3-fold rotation center can be at any sublattice in a unit cell, the vibration of the rotation-center sublattice must be an eigenstate of the rotation operator $\Re(\frac{2\pi}{3},z)$ due to no phase change for the orbital part of it. Therefore for non-degenerate modes, sublattices must do circularly polarized vibration otherwise they are still, which is consistent to our findings in Fig.~\ref{fig1} (a).

The orbital pseudo angular momentum can be obtained from phase change under a 3-fold rotation, which is shown in Fig.~\ref{fig1} (b). We can obtain $l^o_A=\tau$ and $l^o_B=-\tau$, $\tau=\pm1$ labels the two valleys $\bm K$ and $\bm K'$. Since phonon state must be an eigenstate of the rotation operator, pseudo angular momentum of phonon equals $l_{\rm ph}=l^s_A+l^o_A=l^s_B+l^o_B$ if both sublattices are vibrating; if one is still, it is decided by the other vibrating sublattice. Therefore we can obtain phonon pseudo angular momentum as listed in Fig.~\ref{fig1} (c).

The chiral valley phonons can be observed in graphene systems with introducing isotope doping (please see Supplementary information Sec. III for detailed calculation), where the three possibilities of phonon spins are verified. The out-of-plane phonon modes will not play a role in phonon polarization along the direction perpendicular to the lattice plane.
\begin{figure}[t]
\includegraphics[width=3.4 in,  angle=0]{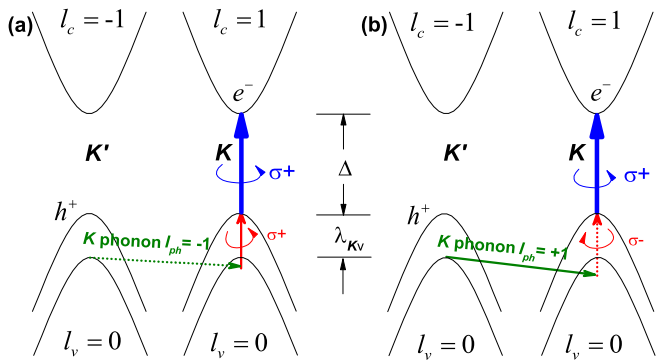}%
\caption{\label{fig2}\textbf{$\bm K$-valley  phonons emitted in hole intervalley scattering in MoS$_2$.} The exciton is excited by a right polarized photon with energy ($\Delta$) at $\bm K$ valley. (a) By absorbing a stimulated right-handed photon with energy ($\lambda_{\bm K v}+ \hbar \omega$), the excited hole in valence band is scattered to the other valley $\bm K'$ by emitting a $\bm K$-valley phonon with energy $\hbar \omega$ and pseudo angular momentum $l_{\rm ph}=-1$. (b) A stimulated left-handed photon is absorbed and a phonon with $l_{\rm ph}=1$ is emitted. The pseudo angular momenta of electrons in conduction band and valence band ($l_{v(c)}=\pm1,0$) are also marked.  }
\end{figure}

\vspace{6pt}
\noindent \textbf {Selection rules.} It is well known that in graphene the double resonance D-peak in Raman spectrum is related to phonon modes in the vicinity of the $\bm K$ point \cite{saito02,malard09}. For electrons with zero moment along the normal direction of the plane ($z$), the state is invariant under a 3-fold discrete rotation, the pseudo angular momentum is decided by the orbits on sublattice A or B.  Based on the lattice structure in Fig.~\ref{fig1} (b), if we assume the valence band corresponds to the orbit on sublattice A and the conduction band corresponds to the orbit on sublattice B, we can obtain all the pseudo angular momenta $l_{c(v)} =\mp \tau $.  Therefore one can expect an azimuthal selection rule $l_c-l_v =\pm \tau$ for interband transition by photons with right ($\sigma+$) or left ($\sigma-$) circular polarization.

A photon excited electron in one valley can decay to another valley by emitting a valley phonon through electron phonon interaction. Due to circularly polarized motion of sublattices at valleys, one can obtain a nonzero ionic magnetic moment, which can provide a possible mechanism for the electron phonon interaction. In the intervalley scattering by phonon the whole system has 3-fold discrete rotation symmetry, thus we can expect a selection rule from the conservation of pseudo angular momentum, that is $l_{c(v)}(\bm K)-l_{c(v)}(\bm K')=\pm1$ by emitting a circularly polarized valley phonon ($l_{\rm ph}=\pm1$), where momentum and energy conservations are also applied.  Since $1+1=-1$ and $\bm K+\bm K=\bm K'$ (or $\bm K'+\bm K'=\bm K$), a phonon with a pseudo angular momentum equal to that of the electron in the same valley. Thus we can expect an valley phonon with a specific pseudo angular momentum can be created. Due to the pseudo angular momentum of valley phonons are different, we can observe a circularly polarized infrared spectrum during the valley phonon interband scattering (please see Sec. IV of the Supplementary information for the detailed discussion).

Due to spin-orbit coupling, the band structure of transition-metal dichalcogenides will have spin splitting $\lambda_{\bm K}$ at ${\bm K}$ valleys, e.g. for highest valence band at valleys MoS$_2$ has a 150 meV splitting, while for the lowest conductance band  only a negligible 3 meV splitting \cite{wang14}. Thus we need consider the splitting $\lambda_{\bm K}$ for energy conserving in electron intervalley scattering where the electronic spin keeps fixed. Furthermore, with a photon absorbtion or emission  we can expect to observe intervalley electron scattering at valley centers involving a valley center phonon, where the selection rules imply $\Delta l_{\rm el} = \pm l_{\rm ph} \pm l_{\rm photon}$ and $\lambda_{\bm K}=\pm \hbar \omega_{\rm ph} \pm \hbar \omega_{\rm photon}$, where '$+$' means emission and '$-$' means absorbtion.

Recently experiments on valley polarization have been reported on some monolayer transition-metal dichalcogenides by optical pumping. In monolayer MoS$_2$, Cao, \emph{et al.} found that the pseudo angular momenta of electrons for valence and conduction bands are quite different from those in gapped graphene. They found valleys in the conduction band have $\pm1$ pseudo angular momentum, while both valleys in the valence band have $0$ as shown in Fig.~\ref{fig2}, thus they found the absorption of circularly polarized photons at valleys.

In Fig.~\ref{fig2}, with spin-orbit coupling  MoS$_2$ has a bandgap of $\Delta=1.65 $eV  and a spin-splitting for the highest valence band at $\bm K$ valley  $\lambda_{\bm K v}=150$ meV  \cite{wang14}. Through a right-handed polarized photon a pair of exciton are excited at $\bm K$ valley, where the blue lines correspond to the absorption of a right-handed photon with energy $\Delta$.  Since the excited electron is in the valley center, which cannot be scattered to another valley through emitting a phonon. However, due to the large spin-splitting of valence band the hole can be scattered to another valley by a circularly polarized photon and phonon by keeping spin fixed. Using the Quantum-Espresso code \cite{gian09}, we obtain phonons for all bands at valleys as shown in Table \ref{table1} (please see Supplementary information Sec. V for detailed calculation). At $\bm K$ valley, phonons with energy 14.4 and 48.0 meV have pseudo angular momentum of $l_{\rm ph} =-1$  while the mode 48.0 meV phonon has a little polarization, and phonons with energy 21.5, 40.0 and 45.1 meV have pseudo angular momentum of $l_{\rm ph} =1$.

\begin{table}[ht]
\caption{\textbf{Chiral phonons in $\bm K$ Valley of MoS$_2$.}  Monolayer MoS$_2$ have 9 modes ($n$) of phonon with energy $\hbar \omega$ (meV), Mo and S are located in A and B respectively as show in Fig. 1 (b) in the main text, thus the  orbital pseudo angular momentum $l^o_{\rm Mo}=1$ and $l^o_{\rm S}=-1$. $s^z_{\rm Mo}$ ($s^z_{\rm S}$), $l^s_{\rm Mo}$ ($l^s_{\rm S}$), and $l_{\rm ph}$  are circular polarizations of Mo (S),  spin pseudo angular momenta of Mo (S),  and phonon pseudo angular momentum. The Mirror symmetry ($M_S$) is relative to the plane of the monolayer of MoS$_2$; it is 1 (-1) if the mode is even (odd) under the mirror symmetry operation. } 
\centering 
\begin{tabular}{c c c c c c c c} 
\hline\hline 
\;$n$ &\;\;$\hbar \omega$\;\; & \;\; $s^z_{\rm Mo}$ \;\;& \;\; $s^z_{\rm S}$\;\; & \;\;$l^s_{\rm Mo}$\; \; &  \; \;$l^s_{\rm S}$\;\;& \;\;$l_{\rm ph}$ \; \;& $M_S$ \;\;\\ [0.5ex] 
\hline 
1 & 14.4 & 0.64 & 0  & 1 & 0 & -1 & 1 \\ 
2 & 21.5 & 0  & -0.34 & 0 & -1 & 1 & -1\\
3 & 31.9 & -0.18 & 0.41 & -1 & 1 & 0 & 1\\
4 & 40.0 & 0 & -0.50 & 0 & -1 & 1 & 1\\
5 & 40.2 & 0 & 0 & - & - & - & -1\\
6 & 41.5 & 0 & 0.50 & 0 & 1 & 0 & 1\\
7 & 45.1 & 0 & -0.40 & 0 & -1 & 1 & -1\\
8 & 48.0 & 0.06 & 0 & 1 & 0 & -1 & 1 \\
9 & 49.2 & -0.34 & 0.33 & -1 & 1 & 0 & 1\\   [1ex] 
\hline 
\end{tabular}
\label{table1} 
\end{table}

Through a right-handed light scanning on the sample, we can observe some resonance peaks on $\lambda_{\bm K v} + \hbar \omega_{\rm ph}$ where stimulated photons are emitted, as shown in Fig. \ref{fig2}(a). The emitted phonon will have a pseudo angular momentum of $l_{\rm ph} =-1$ at $\bm K$ point due to the selection rules at valleys. Thus we can observe a resonance peak of 164.4 meV (another peak of 198.0 meV is not obvious due to the small polarization of phonon).  And for a stimulated left-handed photon in Fig. \ref{fig2}(b), we can only observe one peak of 190.0 meV correspond to phonon mode of 40.0 meV while the other two modes will not be involved since they are odd under mirror operation (see Table I) while in the electron-phonon scattering process the electron wave function keeps even under a mirror operation relative to the x-y plane.  During these two processes, only $\bm K$-valley phonon can be emitted while a chiral phonon at $\bm K'$ can be emitted in the scattering of hole of an exciton at $\bm K'$ valley which can be excited by a left-handed photon with 1.65 eV. After scattered by phonon and photon, the pair of electron and hole locate in different valleys and can have a long life time; its condensation has been theoretically observed very recently \cite{wu15}.  Therefore, a specific chiral phonon at a definite valley can be obtained through a stimulated photon.  With the two-step polarized light shinning on the sample, a large number of valley phonons with definite frequencies can be created.

\begin{figure}[t]
\includegraphics[width=3.4 in, angle=0]{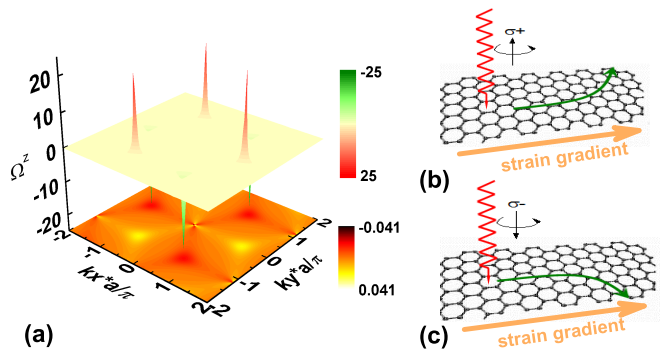}%
\caption{\label{fig3}\textbf{Phonon Berry curvature and valley phonon Hall effect in a honeycomb lattice.} (a) Berry curvature of band 1 (bottom contour plot) and band 2 (top 3D plot). Band 3 (Band 4) has phonon angular momentum opposite to that of band 2 (band 1). (b) ((c))  Schematic of valley phonon Hall effect (Hall current denoted by olive curve arrows) under a strain gradient (orange arrows), where valley phonons are excited by a ray of right-handed or left-handed polarized light (red wave lines). The parameters are the same with those in Fig. 1.  }
\end{figure}

\textbf{Valley phonon Hall effect.} In the presence of an in-plane electric field, an electron will acquire an anomalous velocity proportional to the Berry curvature in the transverse direction \cite{chang96,xiao10}. Recently the electronic valley Hall effect proposed in \cite{xiao07} has been experimentally observed in monolayer MoS$_2$ transistors \cite{mak14} and in graphene superlattices \cite{gor14}.  As discussed above, phonons with definite frequencies at a specific valley can be massively created, thus for valley phonons, if its Berry curvature is nonzero we can also expect to observe valley phonon Hall effect in the presence of an in-plane gradient strain field.  Such valley phonon Hall effect can provide us another way to observe valley phonons.

With the breaking of spatial inversion symmetry we observe nonzero phonon Berry curvature at valleys as shown in Fig.~\ref{fig3} (a) (see Supplementary information Sec. VI  for derivation). Band 2 and band 3 have large Berry curvatures at valleys, while those of band 1 and 4 are small. 
Due to the nonzero phonon Berry curvature, applying a strain gradient $\bm E_{\rm strain}$ along $x$ direction,  phonons excited at a different valley will go to a different  transverse direction since ${\bm v}_{\rm anom} \propto - \bm E_{\rm strain} \times \bm \Omega$ in analogy to electrons. If the photon polarization is reversed, the transverse phonon current would be reversed as shown in Fig.~\ref{fig3} (b) and (c).
\\

We thank Ji Feng and Gang Zhang for helpful discussions. We acknowledge support from  DOE-DMSE (DE-FG03-02ER45958),NBRPC (2012CB-921300), NSFC (91121004), and the Welch Foundation (F-1255).

\end{document}